\documentclass[11pt,twoside]{article}
\usepackage{asp2010}
\resetcounters

\aspvolume{455}
\aspcpryear{2012}
\aspvoltitle{4$^{th}$ {\em Hinode} Science Meeting: Unsolved 
Problems and Recent Insights}
\aspvolauthor{L.~R.~Bellot Rubio, F.~Reale, and M.~Carlsson, eds.}

\begin{document}

\title{The Role of Coronal Hole and Active Region Boundaries in 
Solar Wind Formation}
\author{Louise K.~Harra}
\affil{UCL-Mullard Space Science Laboratory, Holmbury St Mary, Dorking, Surrey, RH5 6NT, UK}

\begin{abstract}
{\em Hinode} observations have provided a new view of outflows from the
Sun. These have been focussed in particular on flows emanating from
the edges of active regions. These flows are long lasting and seem to
exist to some extent in every active region. The flows measured have
values ranging between tens of~km~s$^{-1}$ and 200~km~s$^{-1}$. Various
explanations have been put forward to explain these flows including
reconnection, waves, and compression. Outflows have also been observed
in coronal holes and this review will discuss those as well as the
interaction of coronal holes with active regions. Although outflowing
plasma has been observed in all regions of the Sun from quiet Sun to
active regions, it is not clear how much of this plasma contributes to
the solar wind. I will discuss various attempts to prove that the
outflowing plasma forms part of the solar wind.
\end{abstract}

\section{Introduction}

The slow solar wind is more dynamic and has speeds of a few
hundred~km~s$^{-1}$ whereas the fast solar wind is steady with speeds
of 800~km~s$^{-1}$. The fast solar wind originates from regions where
the magnetic fields lines are open. The slow solar wind is more
complicated with several different scenarios to explain its
origins. These include reconnection between open coronal hole and
closed helmet streamer magnetic field
\citep[e.g.,][]{2004JGRA..10903107C}, reconnection between open flux
and closed flux which displaces the open field lines
\citep[e.g.,][]{2003JGRA..108.1157F}, and reconnection between closed
loops associated with active regions and nearby open
flux. Irrespective of which scenario you choose to explain the slow
solar wind, it is clear that the boundaries of both coronal holes and
active regions are important in the formations and development of both
the slow and fast solar wind.

\begin{figure}
\centerline{\includegraphics[width=0.95\textwidth,clip=]{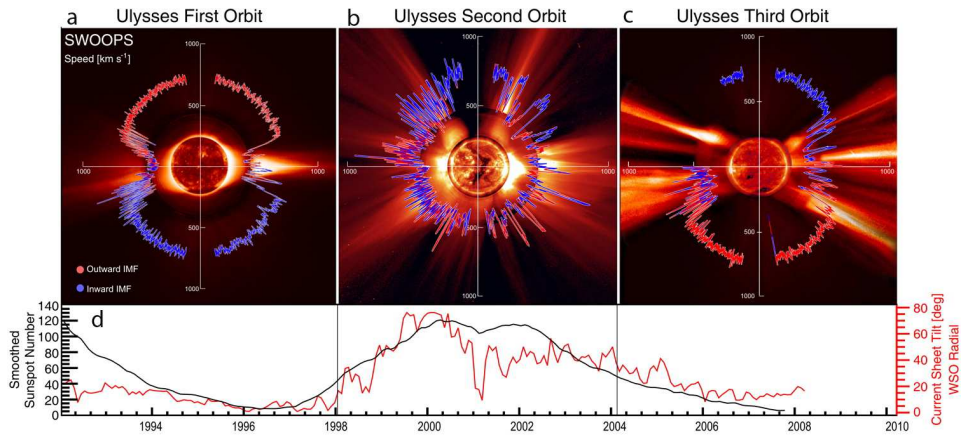}}
\caption{{\em Top:} Polar plots of the solar wind speed from three orbits 
obtained by the Ulysses spacecraft. {\em Bottom:} Sunspot number and
heliospheric current sheet tilt during the same time period. From
\citet{2008GeoRL..3518103M}.  Copyright 2008 American Geophysical
Union. } \label{fig:mccomas}
\end{figure}

\citet{2008GeoRL..3518103M} shows the change in the behaviour of 
the solar corona during the first Ulysses orbit (during a solar
minimum period), the second Ulysses orbit (during a solar maximum), and
the final Ulysses orbit (during solar minimum again) (see 
Figure~\ref{fig:mccomas}). The first orbit shows a 'classic' solar minimum
with the Sun taking the form of a clear dipole with the streamer belt
at the equator. The heliospheric current sheet tilt is very low. The
second orbit shows a 'classic' solar maximum period where the streamer
belt is chaotic and reaches high latitudes, with the polar coronal
holes becoming less clear. The heliospheric current sheet tilt was
high. The final orbit is the most recent solar minimum at the start of
solar cycle 24. During this period the streamer belt was not lying
along the equator and the heliospheric current tilt is higher than the
previous solar minimum. The interactions between open and closed field
drive the changes observed in the solar cycle. In this review we will
discuss some potential sources of the slow and fast solar wind from
active region, coronal hole, and quiet Sun boundaries and discuss
whether these sources do actually form part of the solar wind.

\section{Coronal Holes, Their Boundaries, Their Formation and Larger
 Scale Impact}
 
\begin{figure}[!t] 
\centerline{\includegraphics[width=0.96\textwidth,clip=]{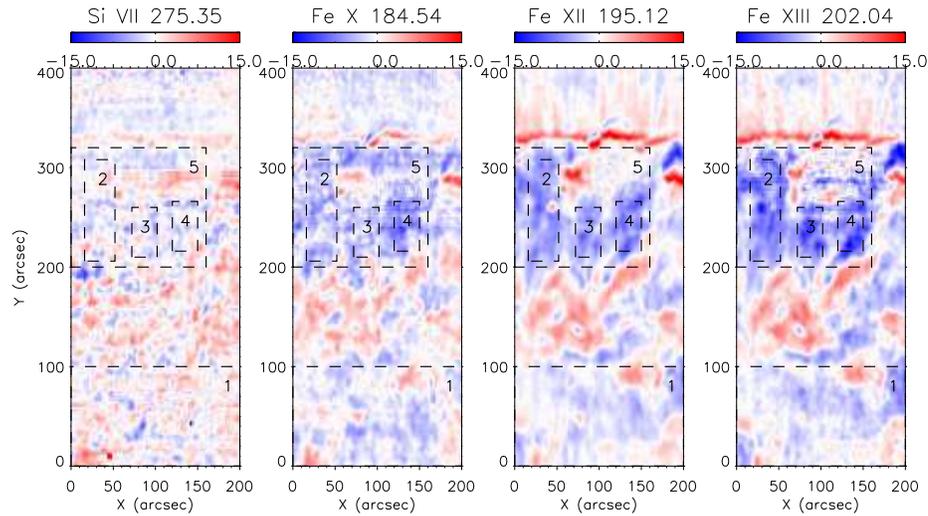}}
\caption{Doppler velocity maps of the north polar coronal
hole from EIS, with blue showing blue-shifted
plasma and red showing red-shifted plasma. The Doppler velocity range
is $\pm 15$~km~s$^{-1}$. From \citet{2010ApJ...709L..88T}. Reproduced by
permission of AAS. } \label{fig:tian} \end{figure}

The fast solar wind is known to emanate from coronal holes where the
magnetic field lines are predominantly open to interplanetary
space. Polar coronal holes are in some senses simpler than equatorial
coronal holes as they have fewer interactions with surroundings that
equatorial corona holes that lie in the activity belt. Hence we begin
by describing polar coronal holes.  Strong magnetic field patches have
been discovered in the polar coronal holes. These have been compared
to the quiet Sun magnetic field by \citet{2010ApJ...719..131I} and
they found that the average area and the total magnetic flux of the
kilo-Gauss magnetic concentrations in the polar region is larger than
those of the quiet Sun. \citet{2010ApJ...709L..88T} studied the
temperature behaviour of the outflows with EIS in a polar coronal hole
and found that the outflow appears to start in the transition
region. Figure~\ref{fig:tian} illustrates outflowing plasma in
different part of the solar atmosphere. The outflow becomes more
prominent with increasing temperature. Significant outflows appear 
as small patches in the transition region, and these patches merge
together as the temperature of the plasma increases. This is
consistent with the concept of the solar wind being guided by
expanding magnetic funnels as described by \citet{2005Sci...308..519T}
where the plasma is accelerated above 5~Mm.
 
\begin{figure} 
\centerline{\includegraphics[width=0.9\textwidth,clip=]{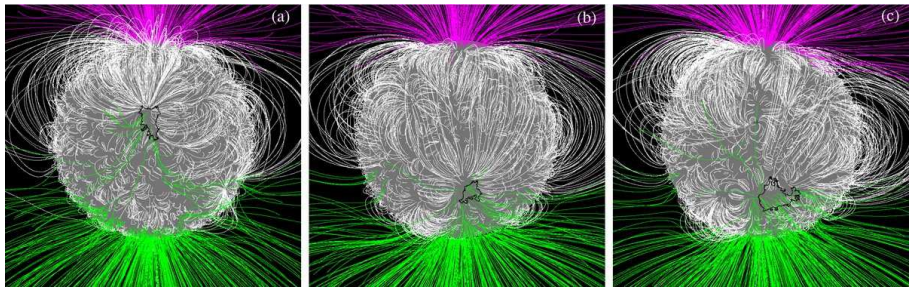}}
\caption{Magnetic field extrapolations highlighting the linkage 
between the three equatorial coronal holes and the north polar coronal
hole during the recent solar minimum period. From
\cite{2010ApJ...714.1672K}, reproduced by permission of AAS.  }
\label{fig:karachik}
\end{figure}

Equatorial coronal holes, on the other hand, have the potential to
interact with the more 'active' Sun. The formation of equatorial
coronal holes and their interaction with their surroundings is
important in terms of understanding the solar
cycle. \cite{2010ApJ...714.1672K} have studied a number of cases where
equatorial coronal holes have formed during active region decay. As
active regions decay, the leading and lagging magnetic polarities
usually exhibit different dissipation rates. This causes an increase
in the magnetic flux imbalance as the region decays, and a coronal
hole will form in the place where the most compact polarity existed in
the active region. Figure~\ref{fig:karachik} shows magnetic field
extrapolations of three examples of coronal holes formed followed
active region decay. All the regions studied were observed during the
end of solar cycle 23 and beginning of solar cycle 24.  The equatorial
coronal holes had their magnetic field ending in the north polar
coronal hole, regardless of which hemisphere the coronal hole was
located. This indicates the flux imbalance between the north and south
coronal hole during this unusual solar minimum. The behaviour of the
coronal holes is suggestive of the change from toroidal to poloidal
magnetic field that is required to transition from one solar cycle to
another, as is described by the Babcock-Leighton model.
 
We have discussed equatorial and polar coronal holes---but is there a
difference in their characteristics quantitatively?
\citet{2009SoPh..255..119R} analysed the Doppler velocities and line
widths in both coronal holes at the poles and equator and compared
these values with quiet Sun measurements. They found that the coronal
holes have larger blue-shifts and increased line widths than quiet Sun
regions, which would be expected as the magnetic field lines are
expected to be dominantly open in coronal holes, allowing faster
outflow. It was also found that polar coronal holes have larger line
width relative to an equatorial coronal hole. It is also possible that
the reason for this is that the equatorial coronal holes are
surrounded by closed magnetic field and will have interactions with
this.
   
\begin{figure}[!t] 
\centerline{\includegraphics[width=0.9\textwidth,bb=20 0 470 340]{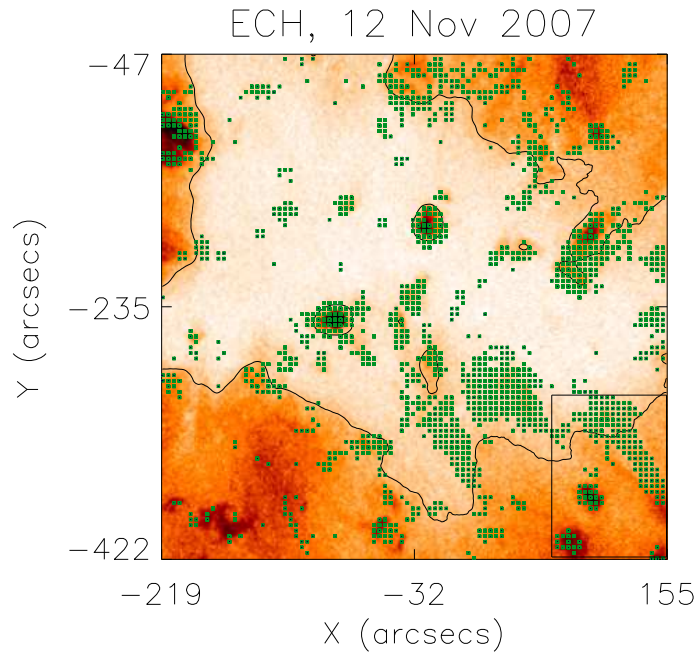}}
\caption{An equatorial coronal hole observed by TRACE. The coronal hole 
boundary is highlighted in black. Positions of identified brightenings
are overlaid in green. From \cite{2010A&A...516A..50S}. Reproduced
with permission \copyright ESO. } \label{fig:maria} \end{figure}

\cite{2010A&A...516A..50S} studied the interactions between 
equatorial coronal holes and the surrounding quiet
Sun. Figure~\ref{fig:maria} shows an example where the positions of
brightenings are overlaid. The brightenings were each analysed to
search for evidence of outflow such as expanding loops or collimated
flows. Only 30\% of the brightenings in the quiet Sun showed evidence
of outflowing plasma whereas 70\% of the brightenings in the coronal
hole showed outflowing plasma. The brightenings are often concentrated
at the coronal hole-quiet Sun boundary, the cause of which is likely
to be interchange reconnection between open and closed magnetic field
lines. This process will result in strong plasma outflows that can
form part of the solar wind.
 
Coronal holes, as well as interacting with quiet Sun and other coronal
holes, can also interact with active
regions. \cite{2010SoPh..263..135Y} have studied this interaction
using Yohkoh data to analyse how large-scale trans-equatorial loops
are formed.

As a new active region emerges it interacts with the polar coronal
hole through interchange reconnection. The new loops that are formed
due to this reconnection cause the coronal hole boundary to slowly
retreat as the active region continues to
emerge. \citet{2010SoPh..263..135Y} believe that this boundary
interaction is important in the formation of trans-equatorial loops
(TELs). The magnetic field between the coronal hole and active region
begins as semi-open. As these fields interact with the emerging active
region, they form large-scale loop structures. The bright TEL appears
through chromospheric evaporation process which is accompanied by the
eruption of the loop structure.


The large scale structure of the atmosphere is driven to some extent
by coronal holes and their interaction with the surroundings. In the
next section we will divert our attention to active regions and their
boundaries with quiet Sun and coronal holes.

\section{Boundaries of Active Regions with Coronal Holes and Quiet Sun}
 
\begin{figure}[!t]
\centerline{\includegraphics[width=.9\textwidth,clip=]{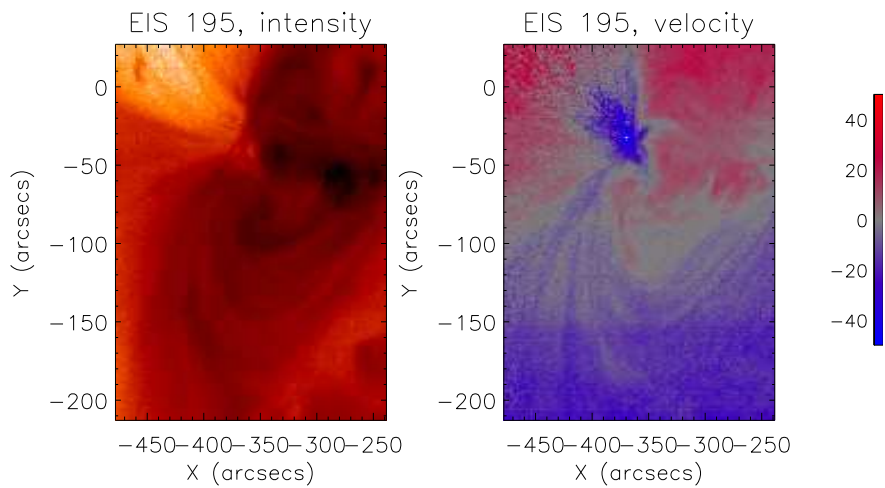}}
\caption{{\em Left:} Intensity image of an active region observed by 
{\em Hinode} EIS in the \ion{Fe}{xii} coronal emission line. {\em
Right:} Doppler velocity data, showing strong outflows as blue. }
\label{fig:harra} \end{figure}

\begin{figure}[!t] 
\centerline{\includegraphics[width=0.72\textwidth,clip=]{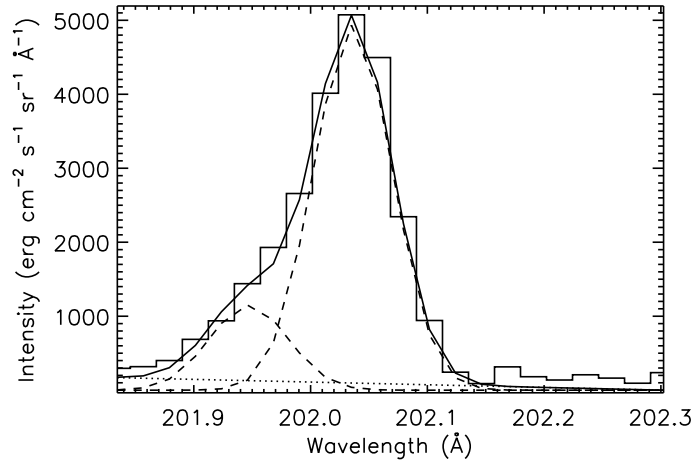}}
\vspace*{-1em}
\caption{The spectral line profile of the \ion{Fe}{xiii} emission line 
observed with EIS at the edge of an active region. There is a very
clear blue component. A double Gaussian profile is used to fit both
components. From Bryans et al. (2010). Reproduced by permission of
AAS. } \label{fig:bryans} \end{figure}

It is clear that active regions can provide an input to the slow solar
wind, but it is not obvious exactly what the source of this is.
Active regions are observed to continuously expand
\citep[e.g.,][]{1992PASJ...44L.155U}. {\em Hinode} observations are now
allowing us to consistently measure outflow at the edges of active
regions (Figure~\ref{fig:harra}). This outflow is observed as steady
streams of outflow in imaging data \citep[see][]{2007Sci...318.1585S}.
Using density measurements they estimated that $\approx$ 25\% of the
slow solar wind could come from such outflowing regions. Spectroscopic
data from EIS shows that there are strong outflows persistently at the
edges of the active region \citep[e.g.,][]{2008ApJ...676L.147H} with
main line profile showing bulk shifts of up to 50~km~s$^{-1}$.
Additional blue-shifted components have been also been observed as
show in Figure~\ref{fig:bryans}.

There are a number of explanations that have been put forward to
explain the outflows. Turbulence at the footpoints was initially
suggested by \citet{2008ApJ...678L..67H}, who found a clear
relationship between the enhanced line widths observed and the Doppler
velocities which suggests unresolved flows are
seen. \citet{2008ApJ...676L.147H} used magnetic field extrapolations
and found that the region of strong outflow was a site of reconnection
with another remote bipole which would allow field lines to be
opened. \citet{2010SoPh..261..253M} found through 2.5D
magnetohydrodynamic simulations of a flux tube emerging into a coronal
hole that the flows could be replicated. Reactionary forces are
generated in the coronal hole magnetic field as the active region
loops expand, which accelerates the plasma.
\citet{2009ApJ...705..926B} determined the location of
quasi-separatrix layers (QSLs), which are the regions that show strong
gradients in the magnetic connectivity. The strongest QSLs are found
to be at the edges of the active regions where the outflows are most
enhanced. This shows that the outflow regions are potential locations
of magnetic reconnection.
 
\begin{figure}[p]
\vspace*{-.5em}
\centerline{\includegraphics[height=.9\textheight]{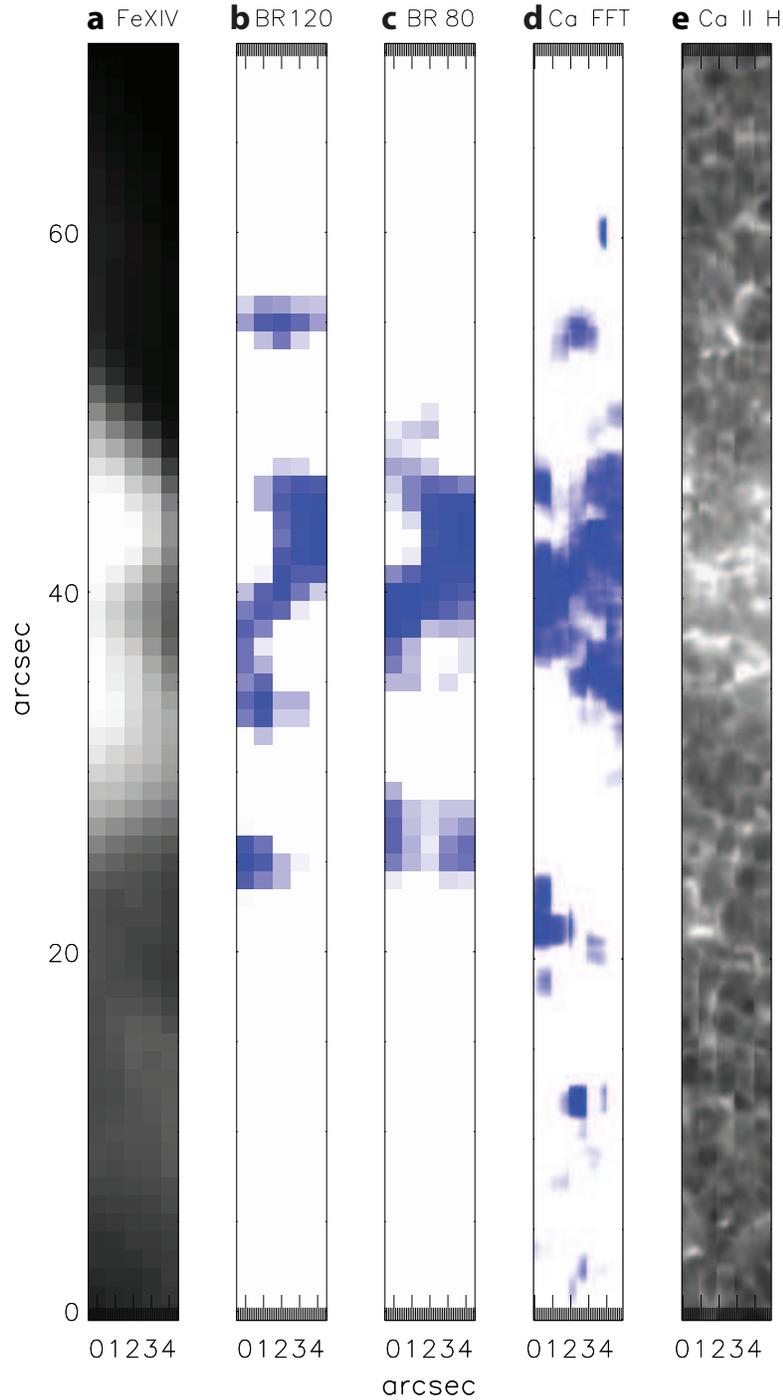}}
\vspace*{-.5em}
\caption{{\em (a)} Intensity of the \ion{Fe}{xiv} line. {\em (b)} Blue-red 
asymmetry centred on 120~km~s$^{-1}$. {\em (c)} Blue-red asymmetry
centred on 80~km~s$^{-1}$, {\em (d)} Chromospheric data from
\ion{Ca}{ii}~H passed through a temporal high-pass Fourier filtering
to show the upper chromospheric dynamics only. {\em (e)} \ion{Ca}{ii}~H
intensity data. From \citet{2009ApJ...701L...1D}. Reproduced by
permission of AAS. }
\label{fig:bart} \end{figure}

\begin{figure}[!t] 
\centerline{\includegraphics[width=0.95\textwidth,clip=]{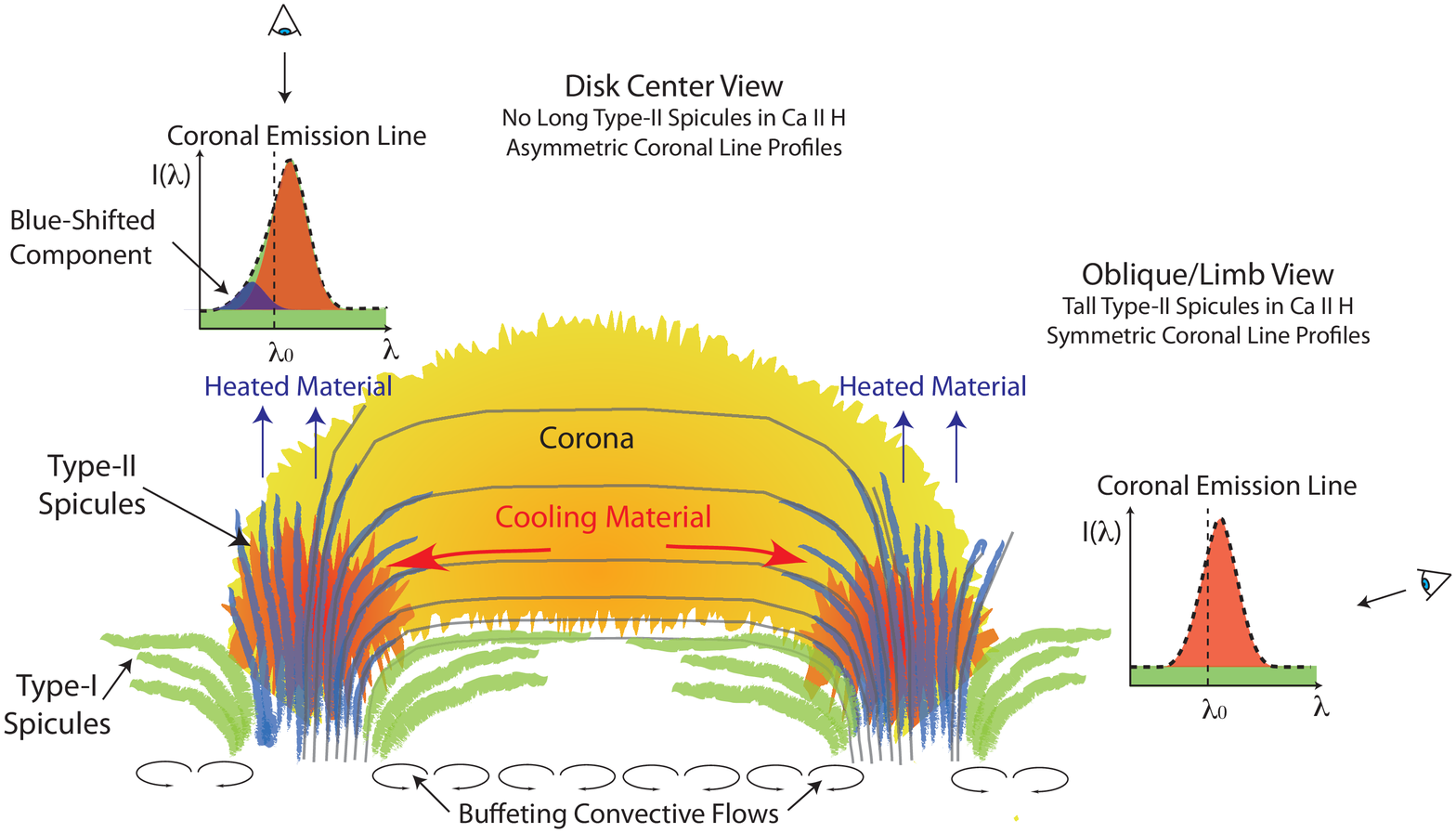}}
\caption{A cartoon describing the possibility for the spicular material 
to be heated to coronal temperature. This would show as a strong but
weak blue shifted component in the corona. From
\citet{2009ApJ...701L...1D}. Reproduced by permission of AAS.}
\label{fig:bart_cartoon} 
\end{figure}

The outflows described above are seen in the corona. The unresolved
outflows have been analysed by using a blue-red asymmetry method which
measures the relative intensities in the red and blue wing of a line
profile. This is a way to measure the weak but strongly blue-shifted
component in the line profile. \citet{2009ApJ...701L...1D} studied the
chromospheric link to the outflows by looking at this blue-red
asymmetry through the atmosphere and found this behaviour to be
consistent throughout the atmosphere. Indeed the patches of strong
blue-red asymmetry seem consistently related to chromospheric
brightenings as seen in Figure~\ref{fig:bart}. Their model to explain
the outflows intimately relates them to chromospheric spicules (see
Figure~\ref{fig:bart_cartoon}). They estimate that only a few percent
of spicules need to be heated to coronal temperatures to heat the
corona.

The chromospheric linkage is not just seen in the line profile
asymmetry but also in the intermittency. \citet{2010A&A...516A..14H}
studied the dynamic behaviour of outflows in different temperature
regimes. They found that the outflows occur 17 times within 5 hours,
and are seen both in X-rays and the EUV with speeds exceeding
200~km~s$^{-1}$. They found chromospheric jets at the root of each
strand that was producing the outflows.

\begin{figure}[t] 
\centerline{\includegraphics[width=0.85\textwidth,clip=]{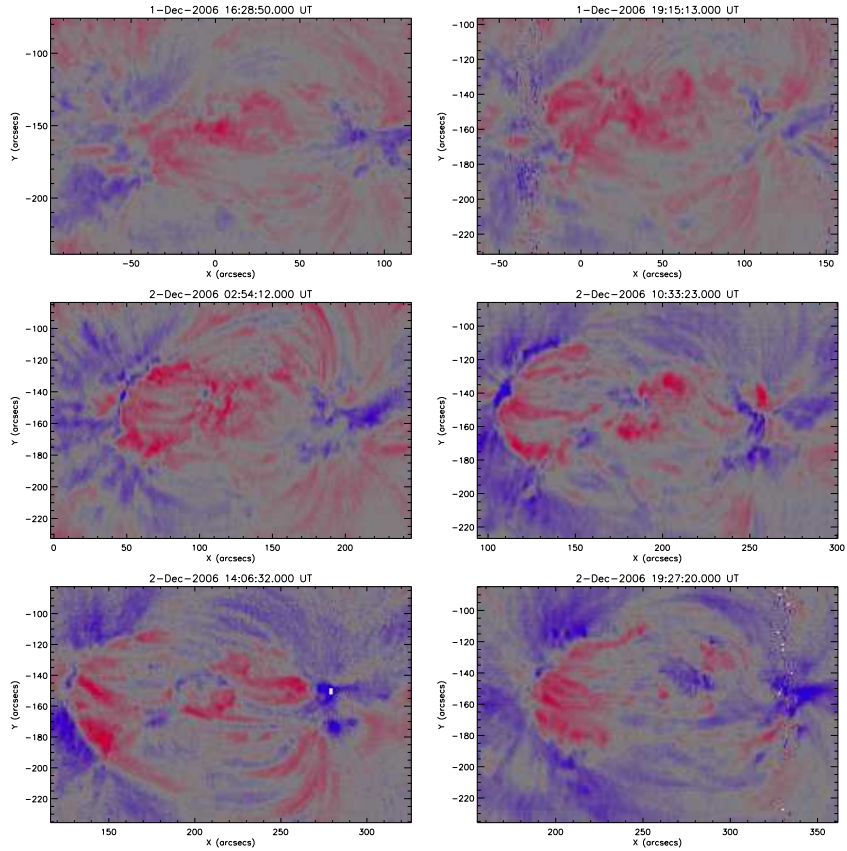}}
\caption{A time series of Doppler velocity images in an active region 
derived from the \ion{Fe}{xiii} emission line. New flux emergence
occurred in the west part of the active region. From
\citet{2010SoPh..263..105H}. }
\label{fig:ef} \end{figure}

The triggering mechanism for the outflows is as yet not
clear. \citet{2010SoPh..263..105H} have analysed an active region
during additional flux emergence into one of the bipoles. The new flux
emergence triggers a reconfiguration of the active region, as would be
expected. One of the responses to the new flux emergence was that
enhanced outflowing material appeared. This was not just on the side
of the new bipole that was most favourable for reconnection, but also
appeared strongly as a clear 'ring' of enhanced Doppler velocities of
the east side of the active region which did not have favourable
magnetic field orientation for reconnection. This enhanced Doppler
velocity lasted for many hours (see Figure~\ref{fig:ef}). We know that
flux emergence exists on all scales from granular scales to active
region scales, and this could also drive the outflows.
 
In this section we concentrated on recent results of outflowing plasma
in active regions. The active regions studied have different
neighbours, be it coronal holes, another active region or quiet
Sun. These interactions will affect the contribution to the solar
wind. A systematic statistical study is needed to progress
understanding further.
 
 \section{Does Any of the Outflowing Plasma Form Part of the Solar Wind?}

Although outflows are seen persistently at the edges of active
regions, and it has been suggested that they could form part of the
solar wind, it is not trivial to prove this. Del Zanna et al. (2010)
have studied noise storms, which are seen at higher altitudes in the
corona and indicate a beam of electrons. They found that the radio
noise storms are persistently seen and are located above the active
region outflows.  They argue for a reconnection scenario that is
pressure driven and continues because of the expansion of the active
region. The outflows are then linked to large-scale open separatrix
field lines and the radio noise storms, and map the electron beams
propagating in the high corona above the active regions. Since the
densities of the active region and nearby coronal hole in their
example are so different, after reconnection occurs, the section
belonging to the former active region will have a much higher density
than the coronal hole. This will lead to the launch of a rarefraction
wave.

\begin{figure}[t]
\centerline{\includegraphics[width=.83\textwidth,clip=]{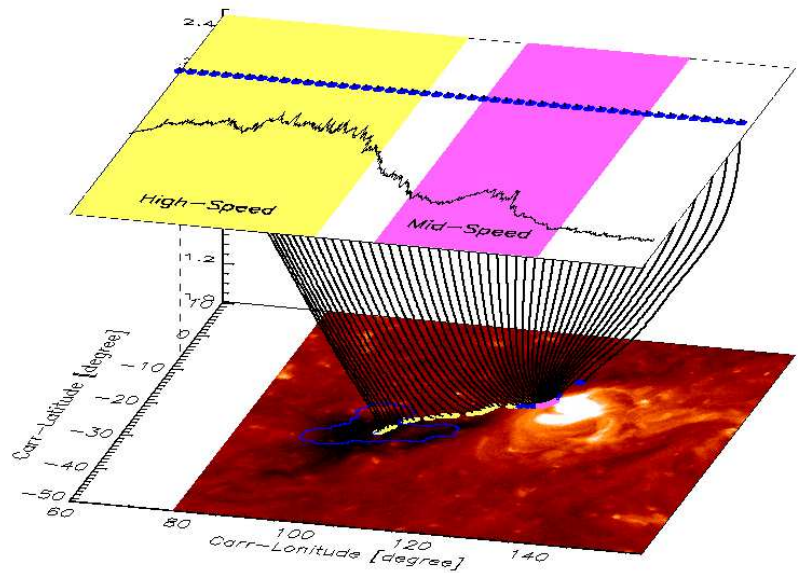}}
\caption{The lower images shows the coronal data with the blue 
contour outlining the coronal hole detected by Kitt Peak. The magnetic
field is extrapolated to 2.5 solar radii. The upper image shows the
solar wind speed measured at 1 AU by ACE. From \citet{2010A&A...516A..14H}. 
Reproduced with permission \copyright ESO. }
\label{fig:he} \end{figure}

To track the outflow beyond the electron beams observed in the radio
wave emission, in-situ data from the ACE spacecraft needs to be
used. \citet{2010A&A...516A..14H} analysed the ACE data for the period
during February 2007 and combined it with solar data. This example was
for an active region close to a coronal hole, so the coronal hole is
easily seen in the ACE data as a high speed stream (see
Figure~\ref{fig:he}). The solar wind stream emanating from the active
region flows has an intermediate velocity.

\begin{figure}[t] 
\centerline{\includegraphics[width=0.95\textwidth,bb=0 0 540 500,clip]{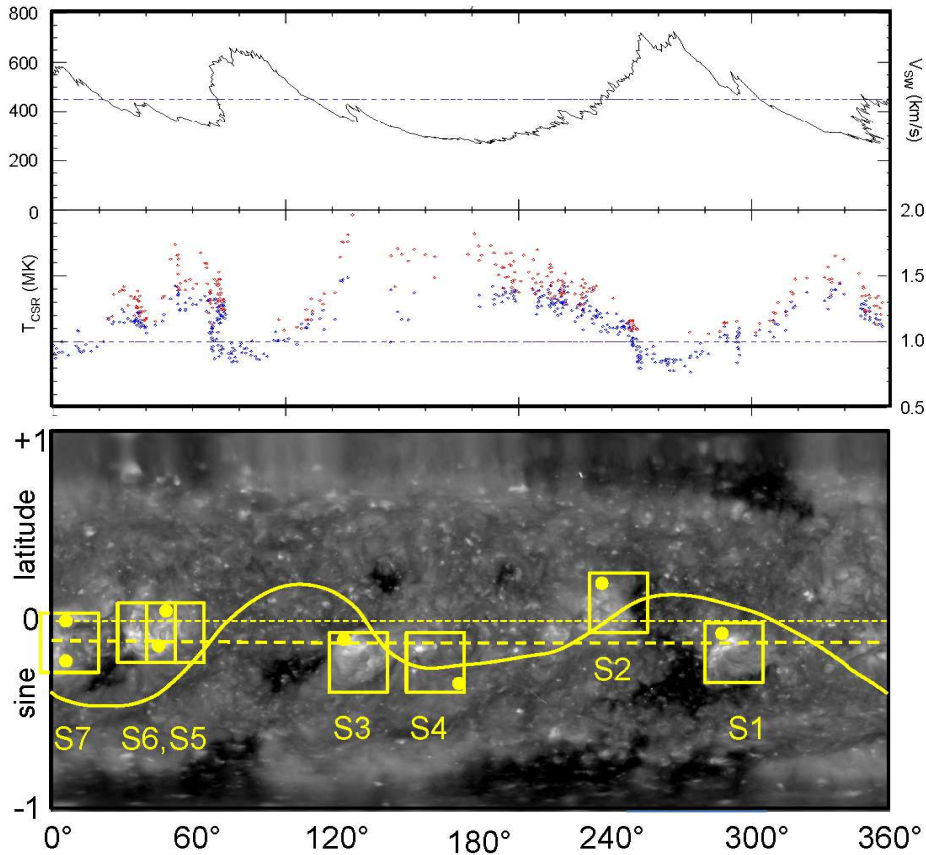}}
\caption{{\em Top:} Wind speed measured by ACE which has been 
back-mapped. {\em Middle:} Oxygen composition in blue. {\em Bottom:}
EUVI synoptic map with the locations of the different
regions studied highlighted by yellow boxes and identified by a
label. Courtesy A.~Fazakerley.}  \label{fig:faz} \end{figure}

A longer term analysis was carried out to study the solar wind related
to active region outflows over a Carrington
rotation. Figure~\ref{fig:faz} illustrates the back-mapped solar wind
and the composition data measured from ACE along with the EUV image of
the corona from the EUV Imager on board STEREO (EUVI). Seven regions
are labeled on the synoptic image. Each of these was observed by EIS
on {\em Hinode} and the approximate location of the strongest outflow
highlighted in the figure. The solar wind speeds related to the active
regions are intermediate speeds, with those active regions located
next to a coronal hole (S1, S3, and S5/S6) showing the most
significant increase in wind speed. The ion composition shows a
confusing picture with values for the 'active region' wind ranging
from similar to coronal hole composition to similar to active region
composition.

In the in-situ data the slow wind plasma typically shows a greater
proportion of elements of low First Ionisation Potential ($<$ 10 eV)
compared to fast wind \citep[e.g.,][]{1999ApJ...521..859S}.  This is
interpreted as indicating a different source for slow wind, in which
electron temperatures are higher, and that slow wind originates on
closed loops.  \cite{2006ApJ...646.1275K} found from analysis of the
Sun at 1.64 solar radii that the source of the slow solar wind seemed
to be predominantly the coronal hole and active region
boundary. Another means to determine whether the outflowing plasma
seen from active regions on the Sun actually forms part of the slow
solar wind is to study the composition at the solar disk.
\citet{2011ApJ...727L..13B}  have studied the relative abundance of 
Si (a low FIP element) and S (a high FIP element) for one active
region over a period of 5 days. They found that Si is enhanced over S
by a factor of 3--4 which is consistent with in situ measurements. This
demonstrates that the plasma flowing from active regions does have the
same composition as that measured in the slow solar wind. This was
however carried out for one active region. The interaction of an
active region with nearby coronal holes and other factors may affect
the composition and dynamics of the outflowing plasma.  Longer term
studies need to be carried out in order to understand the complex
interactions.

\section{Conclusions}

Outflowing plasma is seen in all regions of the Sun on the disk---from
coronal holes, quiet Sun, active region edges, and the interaction
between these three regions.  {\em Hinode} can observed the dynamics
of these regions in great detail. In the future more systematic and
longer term studies are needed to unravel the complexity of the
interactions of one type of phenomena on the Sun to another. In
particular consistent linkage to in-situ data is critical. This will
be of particular interest as the solar cycle slowly progresses to a
maximum, and during the imminent period of polarity reversal.

\acknowledgements \begin{sloppypar} {\em Hinode} is 
a Japanese mission developed and launched by ISAS/JAXA, collaborating
with NAOJ as a domestic partner, NASA and STFC (UK) as international
partners. Scientific operation of the {\em Hinode} mission is
conducted by the {\em Hinode} science team organized at
ISAS/JAXA. This team mainly consists of scientists from institutes in
the partner countries. Support for the post-launch operation is
provided by JAXA and NAOJ (Japan), STFC (UK), NASA (USA), ESA,
and NSC (Norway). \end{sloppypar}

\bibliographystyle{asp2010}
\bibliography{hinode4}

\end{document}